\documentclass[a4paper,prd,preprint,amsmath,amssymb,showpacs,
nofootinbib]{revtex4}
\pdfoutput=1
\usepackage{graphicx}
\begin{document}

\title{Tau longitudinal polarization in $\bar B\to D\tau\bar\nu$ and\\ 
its role in the search for charged Higgs boson} 

\date{July 28, 2010}

\author{Minoru TANAKA}
\email{tanaka@phys.sci.osaka-u.ac.jp}
\author{Ryoutaro WATANABE}
\email{ryoutaro@het.phys.sci.osaka-u.ac.jp}
\affiliation{Department of Physics, Graduate School of Science, 
             Osaka University, Toyonaka, Osaka 560-0043, Japan}

\begin{abstract}
We study the longitudinal polarization of the tau lepton in
$\bar B\to D\tau\bar\nu$ decay. After discussing 
possible sensitivities of $\tau$ decay modes to the $\tau$ polarization,
we examine the effect of charged Higgs
boson on the $\tau$ polarization in $\bar B\to D\tau\bar\nu$. 
We find a relation between the decay rate and 
the $\tau$ polarization, and clarify the role of the $\tau$ polarization
measurement in the search for the charged Higgs boson.
\end{abstract}


\preprint{OU-HET-668-2010}

\maketitle

\section{Introduction}
The quarks and leptons in the third generation are
important clues to new physics beyond the standard model (SM). 
Because of their larger masses their couplings to the electroweak 
symmetry breaking (EWSB) sector are relatively strong, 
and thus their interactions are potentially sensitive to 
new physics that modifies the EWSB sector of the SM.  

The minimal supersymmetric standard model (MSSM) is well-motivated 
and attractive among several candidates of such new physics. 
Its Higgs sector at the tree level corresponds to the two-Higgs-doublet
model (2HDM) of type II and contains a pair of charged Higgs bosons
$H^\pm$ in the physical spectrum. In the 2HDM of type II, 
the first Higgs doublet couples to the down-type quarks and 
the charged leptons, while the second one does to the up-type quarks.
Then, the interaction of the charged Higgs boson  with fermions contains 
terms proportional to $m_f\tan\beta$, where $m_f$ denotes a down-type 
quark mass or a charged lepton one, and $\tan\beta=v_2/v_1$ with 
$v_{1(2)}$ being the vacuum expectation value of the first (second) Higgs 
doublet. Consequently, the charged Higgs contributions to the amplitudes of 
tauonic $B$ decays involve terms proportional to $m_b m_\tau\tan^2\beta$,
and is enhanced if $\tan\beta$ is large.
Several theoretical and experimental studies on tauonic $B$ decays 
have been done motivated by this observation.

The branching fraction of the pure tauonic $B$ decay, 
$B\to\tau\bar\nu$, is measured as 
$(1.7\pm0.6)\times 10^{-4}$ (BABAR) \cite{BABARPT}
and $(1.65^{+0.38+0.35}_{-0.37-0.37})\times 10^{-4}$ (Belle) \cite{BELLEPT}.
Combing them, the Heavy Flavor Averaging Group (HFAG) obtains
$(1.67\pm 0.39)\times 10^{-4}$ (HFAG) \cite{HFAG}.

The theoretical estimation of the decay rate of $B\to\tau\bar\nu$
including the charged Higgs effect is straightforward \cite{HOU93}, 
but suffers from significant uncertainties in the $ub$ element of 
the Cabibbo-Kobayashi-Maskawa (CKM) matrix \cite{CABIBBO,KM} $|V_{ub}|$ 
and the $B$ meson decay constant $f_B$.
Taking the ratio of $B\to\tau\bar\nu$ to $B\to\mu\bar\nu$
does not help, since the lepton universality in the pure leptonic
$B$ decays is not spoiled by the charged Higgs effect \cite{HOU93}. 
Using the branching fraction given by the HFAG, 
$|V_{ub}|=(3.95\pm 0.35)\times 10^{-3}$ \cite{PDG} 
and $f_B=(190\pm 13)\mathrm{MeV}$ \cite{HPQCD}, we obtain the allowed region 
of 95\% C.L. for the charged Higgs parameter as 
$\tan\beta/m_{H^\pm}< 0.11\,\mathrm{GeV}^{-1}$ and 
$ 0.24\,\mathrm{GeV}^{-1}< \tan\beta/m_{H^\pm}< 0.31\,\mathrm{GeV}^{-1}$, 
where $m_{H^\pm}$ denotes the mass of charged Higgs boson.

Semitauonic $B$ decays are more complicated than
the pure tauonic $B$ decay. However, there are several observables
in them besides branching fractions, e.g. decay distributions and 
$\tau$ polarizations. This is one of the reasons that we study
$\bar B\to D\tau\bar\nu$ in the present work. 
Another reason is that it is known to be the most sensitive to the charged
Higgs among several semitauonic $B$ decays studied so far \cite{TANAKA}. 

The charged Higgs effects on the branching fraction, 
the $q^2$ distribution, and $\tau$ polarizations in 
$\bar B\to D\tau\bar\nu$  are investigated theoretically
in the literature 
\cite{TANAKA,GH,GARISTO,TSAI97,WKN,KS,MT,MMT,IKO,CG,KMLAT,NTW,TRINE}. 
The present experimental results on the branching 
fraction of $\bar B\to D\tau\bar\nu$ are given by 
BABAR and Belle collaborations:
\begin{equation}
 \frac{\mathcal{B}(\bar B\to D\tau^-\bar\nu_\tau)}
      {\mathcal{B}(\bar B\to D\ell^-\bar\nu_\ell)}
 =0.416\pm 0.117\pm 0.052\quad\text{BABAR\,\cite{BABARST}}\,,
\end{equation}
and
\begin{eqnarray}
 \frac{\mathcal{B}(B^0\to D^-\tau^+\nu)}
                           {\mathcal{B}(B^0\to D^-\ell^+\nu)}
 &=&0.48^{+0.22+0.06}_{-0.19-0.05}\quad\text{Belle\,\cite{BELLEST1}}\,,\\
 \mathcal{B}(B^+\to \bar D^0\tau^+\nu)
 &=&(0.77\pm 0.22\pm 0.12)\%\,\quad\text{Belle\,\cite{BELLEST2}}\,.
\end{eqnarray}
Averaging them, we obtain the branching-fraction ratio $R$ as
\begin{equation}
R\equiv\frac{\mathcal{B}(\bar B\to D\tau^-\bar\nu_\tau)}
            {\mathcal{B}(\bar B\to D\ell^-\bar\nu_\ell)}
  =0.40\pm 0.08 \quad\mathrm{(average)}\,,
\label{AVST}
\end{equation}
where $\mathcal{B}(B^+\to \bar D^0\ell^+\nu)=(2.15\pm 0.22)\%$
is used \cite{BELLEST1}. A more precise measurement with a few percent 
error in the branching fraction is expected in a super $B$ 
factory \cite{SUPERB}.

The daughter $\tau$'s are identified by successive $\tau$
decays in the experiments: $\tau\to\ell\bar\nu\nu$ ($\ell=e,\mu$) 
is used in Ref.~\onlinecite{BABARST} and Ref.~\onlinecite{BELLEST1}, while 
both $\tau\to\ell\bar\nu\nu$ and $\tau\to\pi\nu$ are used
in Ref.~\onlinecite{BELLEST2}.
The distribution of $\tau$ decay products in $\bar B\to D\tau\bar\nu$ 
is also sensitive to the charged Higgs boson \cite{NTW}. 
It is illustrated in Ref.~\onlinecite{NTW} that the distribution of 
the angle between the momenta of the $D$ meson and the pion in 
$\tau\to\pi\nu$ in the $B$ rest frame depends on the magnitude
and the complex phase of the charged Higgs coupling.
It is also expected that the information on $\tau$ polarizations,
which are affected by the charged Higgs boson 
\cite{TANAKA,GARISTO,TSAI97,WKN}, is encoded in the decay distribution of 
successive $\tau$ decays. 

Effects of $\tau$ polarization on the $\tau$ decay distribution
are well-studied for many $\tau$ production processes, such as
$e^-e^+\to\tau^-\tau^+$ \cite{TSAI71,KST,HPS,PS,KW}, 
$Z^0\to\tau^-\tau^+$ \cite{HMZ,ROUGE,DDDR,BHM}, 
Higgs decays \cite{DPROY,BHM}, 
other heavy particle decays \cite{AAC,NOJIRI,CHKMZ}, 
and the $\nu_\tau$-nucleon scattering \cite{HMY}.
It is shown in the literature that we can decode $\tau$ polarizations 
from appropriate $\tau$ decay distributions in these processes.
We apply a similar method to $\bar B\to D\tau\bar\nu$ in 
the present work.

In this paper, we study the $\tau$ longitudinal polarization in 
$\bar B\to D\tau\bar\nu$ and clarify its role in new physics search
with the main interest in the charged Higgs boson.
It turns out that the $\tau$ longitudinal polarization combined with 
the branching fraction gives us a valuable hint for new physics.
In Sec.~\ref{TAUDECAYS}, we examine possible sensitivities of 
$\tau\to\pi\nu$ and $\tau\to\ell\bar\nu\nu$ to the $\tau$ polarization
in $\bar B\to D\tau\bar\nu$.
Then, we summarize the charged Higgs effects on the decay rate and 
the $\tau$ polarization in Sec.~\ref{HADR}. 
We show our numerical results including a relation between
the decay rate and the $\tau$ polarization in Sec.~\ref{NR}. 
This relation results from a distinctive nature of the charged
Higgs interaction.
Sec.~\ref{CONCLUSIONS} is devoted to conclusions.

\section{Tau polarization and its decay distribution}
\label{TAUDECAYS}
In this section, we illustrate how to extract the $\tau$ polarization in 
$\bar B\to D\tau\bar\nu$ using successive $\tau$ decays and
examine possible sensitivities in experiments.
It is possible to define two distinct and independent $\tau$ polarizations, 
namely the transverse polarization and the longitudinal one.
The transverse polarization is known to be generated by T violating 
interactions and/or final state interactions \cite{GARISTO,TSAI97,WKN}. 
Both interactions are small in the SM. 
While T violating effects induced at one-loop level may 
be sizable in the MSSM. The longitudinal polarization is supposed to be 
sensitive to the chiral structure of the relevant interactions.
The interaction of the charged Higgs boson, which is our main concern 
in the present work, has a different chiral structure from 
that of the $W$ boson. We concentrate on the longitudinal
polarization in the following.

The $\tau$ longitudinal polarization depends on the frame
in which it is defined. We employ the frame in which the spacial
components of the momentum transfer $q^\mu=p_B^\mu-p_D^\mu$ vanish,
where $p_B^\mu$ and $p_D^\mu$ are the four-momenta of 
the parent $\bar B$ meson and the daughter $D$ meson respectively.
We refer to this frame as the $q$ rest frame. Note
that the $q$ rest frame is accessible in the $e^+e^-$ $B$ factories
provided that the tag-side $B$ meson is fully reconstructed.
Incidentally, the $q$ rest frame corresponds to the center of mass
system in $e^-e^+\to\tau^-\tau^+$ in the sense that both are the center
of mass system of the lepton and anti-lepton pair.
In this way, the choice of the $q$ rest frame turns out to be reasonable. 

We use a coordinate system in the $q$ rest frame such that
the direction of the $\bar B$ and $D$ momenta is the $z$ axis,
and the $\tau$ momentum lies in the $x$-$z$ plane.
Then, we parameterize the $\tau$ momentum as
$p_\tau^\mu=E_\tau(1,\beta_\tau\sin\theta_\tau,0,\beta_\tau\cos\theta_\tau)$,
where $E_\tau=(q^2+m_\tau^2)/(2\sqrt{q^2})$,
$\beta_\tau=\sqrt{1-m_\tau^2/E_\tau^2}\,$.
The helicity amplitude of $\bar B\to D\tau\bar\nu$ is
denoted as $\mathcal{M}^{\lambda_\tau}(q^2,\cos\theta_\tau)$, where
${\lambda_\tau}=\pm$ designates the $\tau$ helicity defined in 
the $q$ rest frame and the neutrino helicity is assumed to be negative.
The explicit form of $\mathcal{M}^{\lambda_\tau}$ 
is given in the next section.
The differential decay rate of $\bar B\to D\tau\bar\nu$
for a given $\tau$ helicity ${\lambda_\tau}$ is written as
\begin{equation}
d\Gamma_{\lambda_\tau}=\frac{1}{2m_B}
                \left|\mathcal{M}^{\lambda_\tau}(q^2,\cos\theta_\tau)\right|^2
                d\Phi_3\,,
\label{DGLA}
\end{equation}
where the three-body phase space $d\Phi_3$ is given by
\begin{equation}
 d\Phi_3=\frac{\sqrt{Q_+Q_-}}{256\pi^3 m_B^2}
         \left(1-\frac{m_\tau^2}{q^2}\right)dq^2d\cos\theta_\tau\,,
\end{equation}
and $Q_{\pm}=(m_B\pm m_D)^2-q^2$. The $\tau$ longitudinal polarization
in the $q$ rest frame is defined as
\begin{equation}
P_L(q^2)=\left(\frac{d\Gamma}{dq^2}\right)^{-1}
         \left(\frac{d\Gamma_+}{dq^2}-\frac{d\Gamma_-}{dq^2}\right)\,,
\label{POL}
\end{equation}
where $d\Gamma/dq^2=d\Gamma_+/dq^2+d\Gamma_-/dq^2$, and
we integrate over $\cos\theta_\tau$ since it is difficult to
determine the direction of the $\tau$ momentum at the $B$ factories
in contrast to the case of $Z^0\to\tau^-\tau^+$ at LEP and SLC.
Furthermore, we introduce the average $\tau$ polarization,
\begin{equation}
P_L=\frac{1}{\Gamma}\int dq^2\,\frac{d\Gamma}{dq^2}P_L(q^2)
   =\frac{\Gamma_+-\Gamma_-}{\Gamma}\,,
\label{Q2IPOL}
\end{equation}
where $\Gamma=\Gamma_++\Gamma_-$ is the decay rate of 
$\bar B\to D\tau\bar\nu$.
Though the average polarization holds less information, 
it is still sensitive to the charged Higgs
as we will see below and supposed to be useful for experiments 
with limited statistics.

Measuring the $\tau$ polarization in addition to the decay rate
summed over the $\tau$ helicity, we can separately extract the decay 
rates of fixed $\tau$ helicity. This gives us a clue to new physics 
as we will see later.

The $\tau$ in $\bar B\to D\tau\bar\nu$ is identified by 
$\tau\to\pi\nu$ or $\tau\to\ell\bar\nu\nu$ ($\ell=e,\mu$)
in the present $B$ factory experiments as mentioned above. 
Accordingly, we see how these $\tau$ decay modes work as 
$\tau$ polarization analyzers.
The differential decay rate of the decay chain
$\bar B\to D\tau\bar\nu$ followed by $\tau\to\pi\nu$
($\tau\to\ell\bar\nu\nu$) is written as
\begin{equation}
 \frac{d\Gamma_{\pi(\ell)}}{dq^2d\zeta}=
  \mathcal{B}_{\pi(\ell)}\frac{d\Gamma}{dq^2}\,
  [f(q^2,\zeta)+P_L(q^2)\,g(q^2,\zeta)]
\label{CHAIN}
\end{equation}
where $\zeta=E_{\pi(\ell)}/E_\tau$ with $E_{\pi(\ell)}$ being the
$\pi(\ell)$ energy in the $q$ rest frame and $\mathcal{B}_{\pi(\ell)}$ 
denotes the branching fraction of $\tau\to\pi\nu$ ($\tau\to\ell\bar\nu\nu$).

The functions $f$ and $g$ for $\tau\to\pi\nu$
are well-known and given by
\begin{equation}
f(q^2,\zeta)=1/\beta_\tau\,,\quad g(q^2,\zeta)=(2\zeta-1)/\beta_\tau^2\,,
\end{equation}
where we neglect the pion mass for simplicity, and
the range of $\zeta$ is $(1-\beta_\tau)/2\leq\zeta\leq(1+\beta_\tau)/2$.

As for $\tau\to\ell\bar\nu\nu$, ignoring the $\ell$ mass, 
the decay distribution is described by
\begin{eqnarray}
f(q^2,\zeta)&=&\frac{16}{3}\frac{\zeta^2}{(1-\beta_\tau^2)^3}
           [9(1-\beta_\tau^2)-4(3+\beta_\tau^2)\zeta]\,,\\
g(q^2,\zeta)&=&-\frac{16}{3}\frac{\zeta^2}{(1-\beta_\tau^2)^3}
            \beta_\tau [3(1-\beta_\tau^2)-16\zeta]\,,
\end{eqnarray}
for $0\leq\zeta\leq (1-\beta_\tau)/2$, and
\begin{eqnarray}
\label{ELLFHIGH}
f(q^2,\zeta)&=&\frac{1+\beta_\tau-2\zeta}{3\beta_\tau(1+\beta_\tau)^3}
           [5(1+\beta_\tau)^2+10(1+\beta_\tau)\zeta-16\zeta^2]\,,\\
\label{ELLGHIGH}
g(q^2,\zeta)&=&\frac{1+\beta_\tau-2\zeta}{3\beta_\tau(1+\beta_\tau)^3}
           \frac{1}{\beta_\tau}[(1+\beta_\tau)^2+2(1+\beta_\tau)\zeta
                                -8(1+3\beta_\tau)\zeta^2]\,,
\end{eqnarray}
for $(1-\beta_\tau)/2\leq\zeta\leq (1+\beta_\tau)/2$. 
Eqs.~(\ref{ELLFHIGH}) and (\ref{ELLGHIGH}) reduce to the more familiar 
functions in the collinear limit $\beta_\tau\to 1$, see e.g. 
Ref.~\onlinecite{BHM}.

We can determine $P_L(q^2)$ by measuring the $\zeta$ distribution for 
fixed $q^2$ in Eq.~(\ref{CHAIN}). The statistical uncertainty
of the ideal experiment is given by \cite{ROUGE,DDDR}
\begin{equation}
\delta P_L(q^2)=\frac{1}{\sqrt{N(q^2)}S(q^2)}\,,
\end{equation}
where $N(q^2)$ is the number of signal events for fixed $q^2$
(or in a bin of $q^2$, more practically) and
\begin{equation}
S(q^2)=\left[\int d\zeta\frac{g^2(q^2,\zeta)}
                             {f(q^2,\zeta)+P_L(q^2)g(q^2,\zeta)}
       \right]^{1/2}\,.
\end{equation}
For the average polarization $P_L$ in Eq.~(\ref{Q2IPOL}),
we obtain
\begin{equation}
\delta P_L=\frac{1}{\sqrt{N}\,S}\,,
\end{equation}
where $N$ is the total number of signal events and
the average sensitivity $S$ is given by 
\begin{equation}
S=\left[\frac{1}{\Gamma}\int dq^2\frac{d\Gamma}{dq^2}S^{-2}(q^2)
  \right]^{-1/2}\,.
\end{equation}
Assuming the SM and neglecting the uncertainties
in the form factors discussed in Sec.~\ref{NR},
we obtain $S=0.60$ and $0.23$ for $\tau\to\pi\nu$ and 
$\tau\to\ell\bar\nu\nu$ respectively. These values vary less than 20\% 
even in the presence of charged Higgs boson taking the constraint from 
the branching fraction into account. 

The expected uncertainty in $P_L$ is $\delta P_L\sim 0.4$ 
with $N\sim 100$ for $\tau\to\ell\bar\nu\nu$,
which corresponds to the present experimental status \cite{BABARST,BELLEST1}.
As for $\tau\to\pi\nu$, 
$\delta P_L\sim 0.3/\sqrt{\varepsilon_\pi/\varepsilon_\ell}$, 
is expected in the present experiments, where the branching 
fractions of $\tau\to\pi\nu$ and $\tau\to\ell\bar\nu\nu$ are 
taken into account, and $\varepsilon_{\pi(\ell)}$ represents 
the efficiency of the $\tau\to\pi\nu$ ($\tau\to\ell\bar\nu\nu$) mode.
At the super $B$ factory with integrated luminosity of 
$50\,\mathrm{ab}^{-1}$, $N\sim 2000(3000)$ for $\tau\to\pi\nu$ 
($\tau\to\ell\bar\nu\nu$) is obtained based on the Monte Carlo 
simulation in Ref.~\onlinecite{SUPERB} and thus $\delta P_L\sim 0.04(0.08)$
is expected%
\footnote{We assume that efficiencies of $\tau\to\pi\nu$ and
$\tau\to\rho\nu$ are the same.}.

\section{Helicity amplitudes and decay rates}
\label{HADR}
In the presence of charged Higgs boson, 
both the $W$ boson and the charged Higgs boson contribute to 
the helicity amplitude of $\bar B\to D\tau\bar\nu$. 
We describe their contributions in turn.

The $W$ boson exchange amplitude $\mathcal{M}^{\lambda_\tau}_W$ is written as
\cite{HMW89,HMW90}
\begin{equation}
\mathcal{M}^{\lambda_\tau}_W(q^2,\cos\theta_\tau)=
 \frac{G_F}{\sqrt{2}}V_{cb}
 \sum_{\lambda_W}\eta_{\lambda_W} H_{\lambda_W}L_{\lambda_W}^{\lambda_\tau}\,,
\label{WEX}
\end{equation}
where $V_{cb}$ is the $cb$ element of the CKM matrix, 
$\lambda_W=\pm,0,s$ denotes the virtual
$W$ helicity, and the metric factor $\eta_{\lambda_W}$ is given by 
$\eta_{\pm,0}=1$ and $\eta_s=-1$. The hadronic amplitude 
$H_{\lambda_W}$ that represents the process $\bar B\to DW^*$
is defined by
\begin{equation}
H_{\lambda_W}(q^2)=
 \epsilon_\mu^*(\lambda_W)\langle D(p_D)|
                           \bar c\gamma^\mu(1-\gamma_5) b
                          |\bar B(p_B)\rangle\,,
\label{HAW}
\end{equation}
where $\epsilon_\mu(\lambda_W)$ is the polarization vector of
the virtual $W$ boson. The leptonic amplitude $L_{\lambda_W}^{\lambda_\tau}$
that represents the process $W^*\to \tau\bar\nu_\tau$ 
is defined by
\begin{equation}
L_{\lambda_W}^{\lambda_\tau}(q^2,\cos\theta_\tau)=
 \epsilon_\mu(\lambda_W)\langle\tau(p_\tau,\lambda_\tau)\bar\nu_\tau(p_\nu)|
                         \bar \tau\gamma^\mu(1-\gamma_5)\nu_\tau
                        |0\rangle\,.
\label{LAW}
\end{equation}

Here, we introduce the hadronic form factors $h_\pm(w)$ \cite{NEUBERT91},
\begin{equation}
\langle D(v')|\bar c\gamma^\mu b|\bar B(v)\rangle=
 \sqrt{m_Bm_D}\left[h_+(w)(v+v')^\mu+
                    h_-(w)(v-v')^\mu\right]\,,
\label{HFD}
\end{equation}
where $v^\mu=p_B^\mu/m_B$, $v^{\prime\mu}=p_D^\mu/m_B$ and
$w=v\cdot v'$.
The hadronic amplitudes are written in terms of these form factors:
\begin{eqnarray}
H_\pm(q^2)&=& 0\,,\\
H_0(q^2)&=&\sqrt{m_B m_D}\,
           \frac{1+r}{\sqrt{1-2rw+r^2}}\sqrt{w^2-1}\,V_1(w)\,,\\
H_s(q^2)&=&\sqrt{m_B m_D}\,
           \frac{1-r}{\sqrt{1-2rw+r^2}}(w+1)\,S_1(w)\,,
\end{eqnarray}
where $r=m_D/m_B$, and
\begin{eqnarray}
V_1(w)&=&h_+(w)-\frac{1-r}{1+r}\,h_-(w)\,,\\
S_1(w)&=&h_+(w)-\frac{1+r}{1-r}\,\frac{w-1}{w+1}\,h_-(w)\,.
\end{eqnarray}
In the heavy quark limit (HQL), $h_+(w)$ reduces to the universal form factor
known as the Isgur-Wise function $\xi(w)$ with the normalization $\xi(1)=1$, 
and $h_-(w)$ vanishes \cite{IW}. $V_1(w)$ and $S_1(w)$
also reduce to the Isgur-Wise function in the HQL.

The required leptonic amplitudes are explicitly given as
\begin{eqnarray}
L_0^-(q^2,\cos\theta_\tau)&=&
 -2\sqrt{q^2}\,\sqrt{1-m_\tau^2/q^2}\,\sin\theta_\tau\,,\\
L_0^+(q^2,\cos\theta_\tau)&=&
 2m_\tau\sqrt{1-m_\tau^2/q^2}\,\cos\theta_\tau\,,\\
L_s^-(q^2,\cos\theta_\tau)&=& 0\,,\\
L_s^+(q^2,\cos\theta_\tau)&=& -2m_\tau\sqrt{1-m_\tau^2/q^2}\,.
\end{eqnarray}
Note that the leptonic amplitudes other than $L_0^-$ disappear
for massless leptons, and thus the form factor that appears in
$\bar B\to D\ell\bar\nu$ ($\ell=e,\mu$) is only $V_1$.

The helicity amplitude of the charged Higgs exchange is written as
\cite{TANAKA}
\begin{equation}
 \mathcal{M}^{\lambda_\tau}_H(q^2,\cos\theta_\tau)
 =\frac{G_F}{\sqrt{2}}V_{cb}\,\frac{m_b m_\tau}{m_{H^\pm}^2}t^2_\beta\, 
   H_R\,L^{\lambda_\tau}\,,
\label{HEX}
\end{equation}
where the hadronic amplitude $H_R$ is defined as
\begin{equation}
H_R(q^2)=\langle D(p_D)|\bar c(1+\gamma_5)b|\bar B(p_B)\rangle\,,
\label{HAH}
\end{equation}
and the leptonic amplitude is
\begin{equation}
L^{\lambda_\tau}(q^2,\cos\theta_\tau)=
\langle\tau(p_\tau,\lambda_\tau)\bar\nu_\tau(p_\nu)|
\bar \tau (1-\gamma_5)\nu_\tau|0\rangle\,.
\label{LAH}
\end{equation}
The model-dependent coupling factor $t_\beta$ is given
as $t^2_\beta=\tan^2\beta$ in the 2HDM of type II, while
\begin{equation}
t^2_\beta=\frac{\tan^2\beta}
               {(1+\varepsilon_0\tan\beta)
                (1+\varepsilon_\tau\tan\beta)}
\end{equation}
in the MSSM, where $\varepsilon_0$ and $\varepsilon_\tau$ represent
radiative corrections \cite{IKO,BCRS}. 
Using the equations of motion, we relate the hadronic and the leptonic 
amplitudes of the charged Higgs exchange to those of the $W$ exchange
with $\lambda_W=s$ as
\begin{equation}
m_b H_R=\frac{\sqrt{q^2}}{1-r_m}\,H_s\,,\quad
m_\tau L^{\lambda_\tau}=\sqrt{q^2}\,L_s^{\lambda_\tau}\,,
\label{EOM}
\end{equation}
where $r_m=m_c/m_b$. Note that the charged Higgs contributes only to
the amplitude of $\lambda_\tau=+$ and changes the $\tau$ longitudinal
polarization as well as the branching fraction. 

Substituting the total helicity amplitude
$\mathcal{M}^{\lambda_\tau}=
 \mathcal{M}^{\lambda_\tau}_W+\mathcal{M}^{\lambda_\tau}_H$
into Eq.~(\ref{DGLA}) and integrating over $\cos\theta_\tau$,
we obtain
\begin{equation}
 \frac{d\Gamma_{\lambda_\tau}}{dq^2}=
  \frac{G_F^2|V_{cb}|^2}{128\pi^3 m_B^3}\sqrt{Q_+Q_-}
  \left(1-\frac{m_\tau^2}{q^2}\right)^2 F_{\lambda_\tau}(q^2)\,,
 \label{DGAMMAQ2}
\end{equation}
where
\begin{eqnarray}
F_-(q^2)&=&\frac{2}{3}q^2|H_0(q^2)|^2\,,
\label{FM}\\
F_+(q^2)&=&m_\tau^2\left[\frac{1}{3}|H_0(q^2)|^2+
                         \left|1-\frac{t_\beta^2}{m_H^2}
                               \frac{q^2}{1-z}\right|^2|H_s(q^2)|^2\right]\,.
\end{eqnarray}
We clearly see the negative interference between the charged Higgs
contribution and the standard W boson one as far as $t_\beta^2$ is 
positive. Once the form factors $V_1$ and $S_1$ are given, 
we can evaluate the decay rate $\Gamma_{\lambda_\tau}$ by integrating 
Eq.~(\ref{DGAMMAQ2}) over $q^2$.

As noted above, the charged Higgs boson contributes to the rate of 
$\lambda_\tau=+$, not to $\lambda_\tau=-$. We can test 
this peculiar feature of the charged Higgs boson by measuring 
both the spin-summed decay rate and the $\tau$ longitudinal polarization.

\section{Numerical results}
\label{NR}

\subsection{Form factors}
We employ the following ansatz for $V_1(w)$ \cite{CLN},
\begin{equation}
V_1(w)=V_1(1)\left[1-8\rho_1^2 z+(51.\rho_1^2-10.)z^2
                   -(252.\rho_1^2-84.)z^3\right]\,,
\end{equation}
where $z=(\sqrt{w+1}-\sqrt{2})/(\sqrt{w+1}+\sqrt{2})$.
Since $V_1(w)$ governs $\bar B\to D\ell\bar\nu$
as we mentioned above, the slope parameter $\rho_1^2$ 
is determined by the experimental data of the $q^2$ distribution
in $\bar B\to D\ell\bar\nu$. The recent analysis by 
the HFAG gives $\rho_1^2=1.18\pm 0.04\pm 0.04$ \cite{HFAG}.
To be conservative about the uncertainties in the form factors, 
we combine the above errors linearly in our numerical work.

We parameterize $S_1(w)$ as
\begin{equation}
S_1(w)= \left[1+\Delta(w)\right]V_1(w)\,,
\end{equation}
where $\Delta(w)$ denotes the QCD and $1/m_Q$ corrections.
We estimate the next leading order QCD correction following
Ref.~\onlinecite{NEUBERT94}. In the numerical calculation
of the QCD correction, we use $m_b=4.91\,\mathrm{GeV}$ and
$m_c=1.77\,\mathrm{GeV}$ as the pole masses of the bottom
and charm quarks respectively \cite{XZ}, and $\alpha_s(m_Z)=0.118$
for the running strong coupling \cite{PDG}.
As for the $1/m_Q$ corrections, we take them from Ref.~\onlinecite{CLN}
and use $\bar\Lambda=0.48\,\mathrm{GeV}$ for the mass difference
between a heavy meson and its constituent heavy quark. 

Since the $\overline{\mathrm{MS}}$ scheme
is employed in the calculation of the QCD corrections, 
we should use the same scheme for the quark masses in Eq.~(\ref{EOM}). 
Thus, using the $\overline{\mathrm{MS}}$ masses $\bar m_{b,c}(\mu)$,
$r_m=\bar m_{c}(\mu)/\bar m_b(\mu)$. Note that $r_m$ is
independent of the renormalization scale $\mu$ as it should be.
We use $r_m=0.21$ in the following numerical calculations \cite{XZ}.

The analytic formula of $\Delta(w)$ is rather cumbersome 
and a detailed discussion on it is beyond the scope of this work. 
We only present an approximate expression
\begin{equation}
\Delta(w)=-0.019+0.041(w-1)-0.015(w-1)^2\,,
\end{equation}
which is as good as $3\%$ in the physical range of $w$.
In the following numerical results, we assume $\pm 100\%$ error
in the estimation of $\Delta(w)$, that is, we replace $\Delta(w)$
by $a\Delta(w)$ and vary the uncertainty factor $a$ from 0 to 2. 

\subsection{Decay rate}
Although the effect of charged Higgs on the decay rate
is well studied in the literature, we present our numerical result
to summarize the present status. It is convenient to introduce
a normalized decay rate for each value of $\lambda_\tau$,
\begin{equation}
R_{\lambda_\tau}=\frac{\Gamma_{\lambda_\tau}}{\Gamma_\ell}\,,
\end{equation}
where $\Gamma_\ell=\Gamma_-|_{m_\tau=0}$ is the decay rate of 
$\bar B\to D\ell\bar\nu$. 
We expect that several uncertainties (both theoretical and experimental)
tend to cancel by taking the ratio of the decay rates. 
In particular, the uncertainty in $|V_{cb}|V_1(1)$ disappears in
the theoretical calculation. The branching-fraction ratio defined
in Eq.~(\ref{AVST}) is given by $R=R_++R_-$.

\begin{figure}
\includegraphics[viewport=40 0 800 500,width=38em]{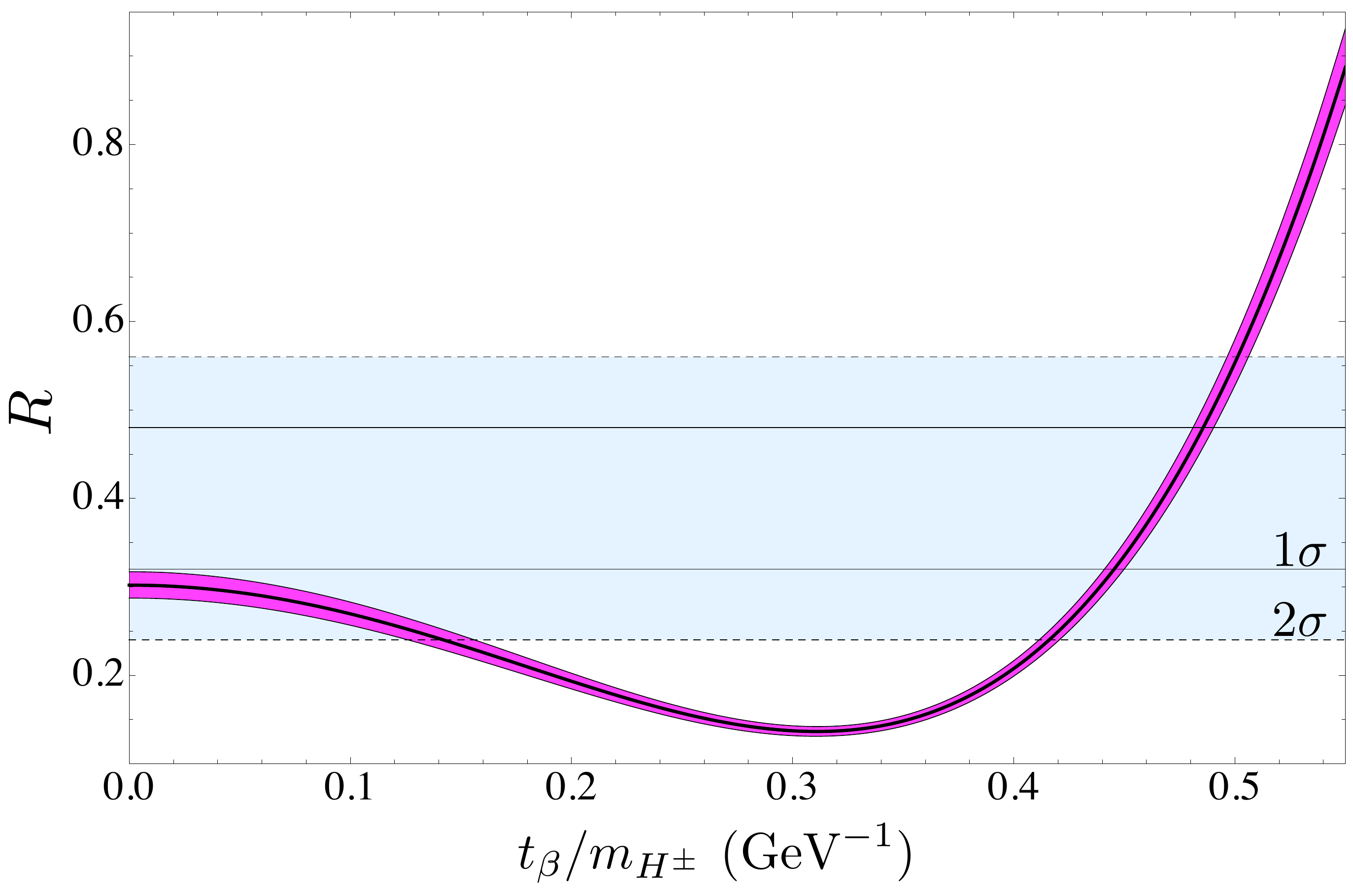} 
\caption{The branching-fraction ratio 
$R=\mathcal{B}(\bar B\to D\tau^-\bar\nu_\tau)/
   \mathcal{B}(\bar B\to D\ell^-\bar\nu_\ell)$ as a function of 
$t_\beta/m_{H^\pm}$. The dark shaded (magenta) band represents 
the theoretical prediction including the uncertainties due to 
$\rho_1^2$ and $a$. The light shaded (light blue) horizontal regions
show the present experimental bounds at $1\sigma$ and $2\sigma$.}
\label{FIG:NDR}
\end{figure}

In Fig.~\ref{FIG:NDR}, we show the branching-fraction ratio $R$
as a function of $t_\beta/m_{H^\pm}$, the control parameter 
of the charged Higgs effect. Hereafter, we take $t_\beta$ to be real 
and positive. The dark shaded (magenta) band represents the theoretical
prediction with the uncertainties in $\rho_1^2$ and $a$. 
The present experimental bounds corresponding to 
Eq.~(\ref{AVST}) are also shown in the figure by the light shaded 
(light blue) horizontal regions. A few comments are in order:
\begin{enumerate}
\item
 The SM prediction is $R|_\mathrm{SM}=0.302\pm 0.015$,
 which does not contradict with those in the literature \cite{CG,NTW}.
\item 
 The present experimental result is consistent with the SM, but
 it seems slightly larger than the SM prediction.
\item
 The allowed regions of $t_\beta/m_{H^\pm}$ are given as 
$t_\beta/m_{H^\pm}<0.14\,\mathrm{GeV}^{-1}$ and 
$0.42\,\mathrm{GeV}^{-1}\,<t_\beta/m_{H^\pm}<0.50\,\mathrm{GeV}^{-1}$.
The latter region, in which the charged Higgs contribution dominates
over the W boson contribution, is practically excluded if combined with
$B^-\to\tau\bar\nu$.
\end{enumerate}

\subsection{Polarization}
\begin{figure}
\includegraphics[viewport=0 0 800 520,clip,width=40em]{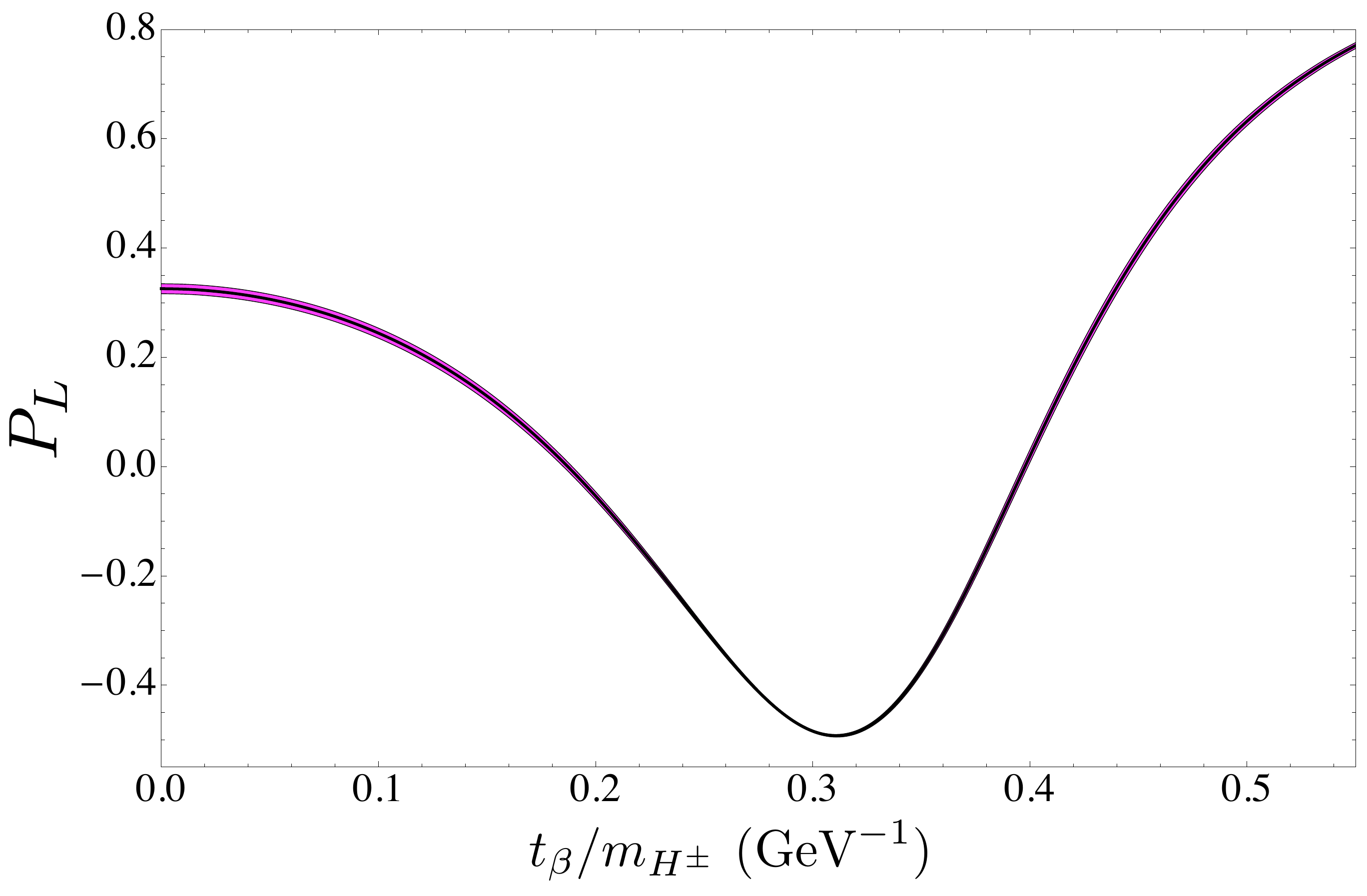} 
\caption{The $\tau$ longitudinal polarization $P_L$ as a function of 
$t_\beta/m_{H^\pm}$. The narrow shaded (magenta) band represents 
the theoretical prediction with the uncertainties due to $\rho_1^2$ and $a$.}
\label{FIG:POL}
\end{figure}

In Fig.~\ref{FIG:POL}, the $\tau$ longitudinal polarization
in the $q$ rest frame is presented as a function of $t_\beta/m_{H^\pm}$.
The width of the band shows the uncertainty in the
theoretical calculation corresponding to $\rho_1^2$ and $a$. 
The SM prediction turns out to be $P_L=0.325\pm 0.009$. 
The theoretical uncertainty is remarkably small and dominated 
by the $a$ factor. The expected statistical uncertainty in the
super $B$ factory is $\delta P_L\sim 0.04$ and larger than
the uncertainty in the SM prediction.

\subsection{Relation between $R$ and $P_L$}
\begin{figure}
\includegraphics[viewport=0 0 800 550,clip,width=40em]{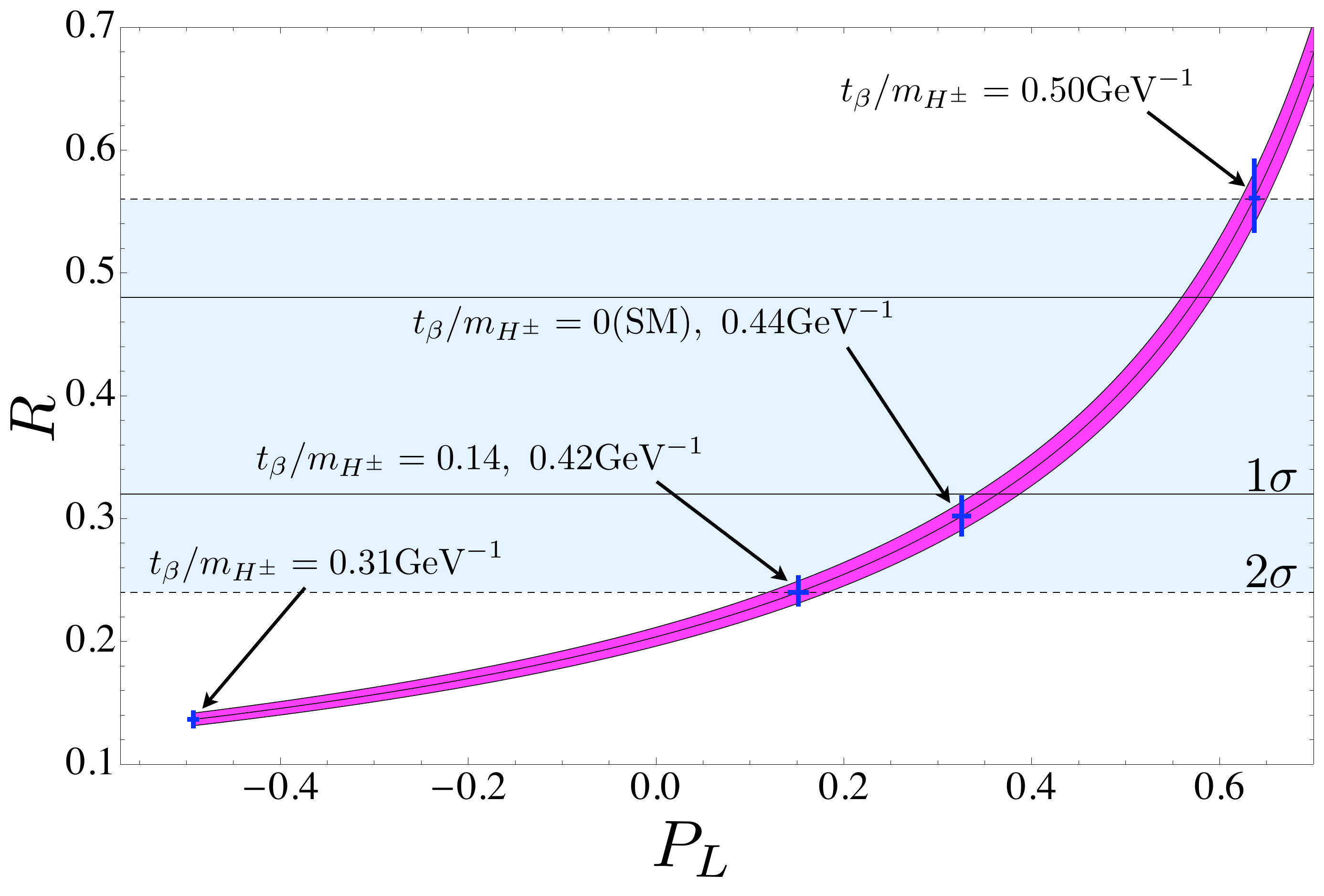} 
\caption{The relation between the branching-fraction ratio $R$ and 
the $\tau$ longitudinal polarization $P_L$. The dark shaded (magenta) 
band represents the relation in Eq.~(\ref{RPLREL}) with the error. 
The present experimental bounds on $R$ at $1\sigma$ and $2\sigma$ are 
also shown by the light shaded (light blue) horizontal regions. 
The (blue) crosses indicate the theoretical predictions on
$R$ and $P_L$ for several values of $t_\beta/m_{H^\pm}$ including
the SM.}
\label{FIG:GAMPL}
\end{figure}

The decay rate and the $\tau$ longitudinal polarization
are independent observables in general. However, as mentioned in
the last paragraph of Sec.~\ref{HADR}, they are related in
the case of the charged Higgs because of the specific chiral
structure of its interaction. It is straightforward to find
\begin{equation}
R(1-P_L)=2R_-=0.204\pm 0.008\,,
\label{RPLREL}
\end{equation}
where $R_-$ is determined only by the standard $W$ boson
contribution as seen in Eq.~(\ref{FM}). We present this relation in
Fig.~\ref{FIG:GAMPL} as the dark shaded (magenta) band with the error.
The light shaded (light blue) horizontal regions show the present
experimental bounds on $R$ at $1\sigma$ and $2\sigma$. 
The present experimental result on $R$ implies $0.15<P_L<0.64$. 
The theoretical predictions on $R$ and $P_L$ for several values of 
$t_\beta/m_{H^\pm}$ including the SM ($t_\beta/m_{H^\pm}=0$) 
are also indicated by the (blue) crosses. The leftmost cross is the turning
point regarding the curve as a trajectory parameterized
by $t_\beta/m_{H^\pm}$. Incidentally, the two-fold ambiguity in
$t_\beta/m_{H^\pm}$ apparently remains. But, it can be solved
combining with $B\to\tau\bar\nu$. 

Eq.~(\ref{RPLREL}) provides a crucial test for the charged Higgs ansatz.
If a set of $R$ and $P_L$ is found out of the dark shaded (magenta)
band in Fig.~\ref{FIG:GAMPL}, it immediately signifies the existence 
of new physics {\em other than} the charged Higgs. 
On the other hand, if one finds it within the band, but away from
the SM prediction, it means that the new physics contributes
to $\Gamma_+$ and not to $\Gamma_-$, and strongly suggests the charged
Higgs.

\section{Conclusions}
\label{CONCLUSIONS}
We have studied the $\tau$ longitudinal polarization in
the $q$ rest frame in $\bar B\to D\tau\bar\nu$.
The $\tau$ polarization is measured through the distribution of
subsequent $\tau$ decays. The expected statistical uncertainty
at the super $B$ factory is $\delta P_L\sim 0.04(0.08)$
for $\tau\to\pi\nu$ ($\tau\to\ell\bar\nu\nu$).

Then, we have examined the effects of the charged
Higgs boson to the decay rate and the $\tau$ polarization
in $\bar B\to D\tau\bar\nu$.
It turns out that the allowed ranges of the charged Higgs parameter 
for the present value of the branching fraction are
$t_\beta/m_{H^\pm}< 0.14\,\mathrm{GeV}^{-1}$ and 
$0.42\,\mathrm{GeV}^{-1}\,<t_\beta/m_{H^\pm}<0.50\,\mathrm{GeV}^{-1}$,
and the uncertainty in the theoretical calculation of the $\tau$ 
polarizations is notably small. 

Furthermore, we have found that the $\tau$ longitudinal
polarization $P_L$ is uniquely related to the branching-fraction ratio $R$ 
in the presence of the charged Higgs effects. This relation reflects 
the specific feature of the charged Higgs interaction.
The present experimental result 
$R= 0.40\pm 0.08$ implies $0.15<P_L<0.64$. 
If a deviation from the SM is found in $R$, the $\tau$ longitudinal
polarization will provide us an important information on the new physics.

\begin{acknowledgments}
The work of MT is supported in part by the Grant-in-Aid for
Science Research, Ministry of Education, Culture, Sports, Science and
Technology, Japan, No. 20244037.
\end{acknowledgments}

\end{document}